\documentclass[12pt]{article}
\usepackage{amsfonts}
\usepackage{amsmath}
\usepackage{amssymb}
\usepackage{fancyheadings}
\usepackage{latexsym}
\usepackage{citesort}

\newcommand{\rf}[1]{(\ref{#1})}

\newcommand{\pt}{\partial}

\def\a{\alpha}
\def\b{\beta}
\def\g{\gamma}

\def\d{\delta}

\def\ep{\varepsilon}

\def\l{\lambda}
\def\m{\mu}
\def\n{\nu}
\def\o{\omega}

\def\s{\sigma}

\def\cN{{\cal N}}

\newcommand{\be}{\begin{equation}}
\newcommand{\ee}{\end{equation}}
\newcommand{\nn}{\nonumber}
\newcommand{\bea}{\begin{eqnarray}}
\newcommand{\eea}{\end{eqnarray}}
\newcommand{\ft}[2]{{\textstyle\frac{#1}{#2}}}
\def\fft#1#2{{\frac{#1}{#2}}}
\def\ul#1{\underline{#1}}
\newcommand{\half}{\frac{1}{2}}

\def\rme{{\rm e}}
\def\rmi{{\rm i}}
\def\rmd{{\rm d}}

\begin{document}

\begin{titlepage}

\font\cmss=cmss10 \font\cmsss=cmss10 at 7pt
\leftline{\tt hep-th/0611111}

\vskip -0.5cm \rightline{\small{\tt KUL-TF-06/28}}

\vskip .7 cm

\hfill
\vspace{18pt}
\begin{center}
{\Large \textbf{The energy and stability of $D$-term strings}}
\end{center}

\vspace{6pt}
\begin{center}
{\large\textsl{Andr{\'e}s Collinucci, Paul Smyth and Antoine Van Proeyen}}

\vspace{25pt}
\textit{Institute for Theoretical Physics, K.U. Leuven,\\ Celestijnenlaan 200D, B-3001 Leuven, Belgium}\\
\end{center}

\vspace{12pt}

\begin{center}
\textbf{Abstract}
\end{center}

\vspace{4pt} {\small \noindent

Cosmic strings derived from string theory, supergravity or any theory of choice should be stable if we hope to observe them. In this paper we consider $D$-term strings in $D=4~,~\cN=1 $ supergravity with a constant Fayet-Iliopoulos term. We show that the positive deficit angle supersymmetric $D$-term string is non-perturbatively stable by using standard Witten-Nester techniques to prove a positive energy theorem. Particular attention is paid to the negative deficit angle $D$-term string, which is known to violate the dominant energy condition. Within the class of string solutions we consider, this violation implies that the negative deficit angle $D$-term string must have a naked pathology and therefore the positive energy theorem we prove does not apply to it. As an interesting aside, we show that the Witten-Nester charge calculates the total gravitational energy of the $D$-term string without the need for a cut-off, which may not have been expected.
}


\vfill
\vskip 5.mm
\hrule width 5.cm
\vskip 2.mm
{\small
\noindent e-mail: andres.collinucci, paul.smyth, antoine.vanproeyen@fys.kuleuven.be}

\end{titlepage}


\section{Introduction}

The resurgence of interest in cosmic strings has been driven by the
realisation that these objects arise naturally in supergravity and string
theory \cite{dkvp,cmp}. A particular class of local cosmic strings  can
be found as solitonic solutions supported by a $D$-term potential in
four-dimensional  $\cN=1$ supergravity with constant Fayet-Iliopoulos
terms \cite{dkvp}. In fact, these $D$-term string solutions were found
previously as the point-like solutions in three-dimensional supergravity
\cite{bbs,e1,e2}, and in the globally supersymmetric Abelian-Higgs theory
in \cite{Davis:1997bs}. A $D$-term string can also be understood as a
D$_{1+q}$-brane wrapping a calibrated $q$-cycle in an internal manifold
of a string theory compactification \cite{dkvp}. The relation between the
two pictures can be established by studying the effective worldvolume
theory of space-filling D-$\overline{\mathrm{D}}$-brane pairs in
Calabi-Yau compactifications of Type II supergravity, where it is
possible to reproduce the $D$-term potential \cite{Douglas:2001ug}. For
example, in flux compactifications, one can use the appropriate
generalised calibrations to explicitly show that the string tension is
set by the Fayet-Iliopoulos term \cite{Martucci:2005ht,martucci}.

Much recent attention has focused on cosmological aspects of string
theory cosmic strings, e.g. string networks \cite{cmp}. However, it is
perhaps surprising to note that the stability of a single, isolated
supersymmetric $D$-term string solution has not been discussed.
Bogomol'nyi bounds for general cosmic strings were constructed some time
ago by Comtet and Gibbons \cite{Comtet:1987wi} and the energy of local
string solutions in current discussions, including the $D$-term strings,
is usually defined using such Bogomol'nyi-type arguments
\cite{dkvp,Achucarro:2005vz,Achucarro:2006ef}. However, as noted in
\cite{Achucarro:2005vz}, a Bogomol'nyi bound does not prove the stability
of such local string solutions, as one is implicitly assuming that the
solutions remain axisymmetric. It is therefore possible that
non-axisymmetric perturbations or string worldvolume perturbations could
lead to instabilities. Bogomol'nyi bounds are useful for displaying
instabilities within such a restricted symmetry class. For instance,
axisymmetric perturbations that grow without bound were found in
\cite{Achucarro:2006ef} for {\it axionic} $D$-term strings by studying
the linearised Bogomol'nyi energy functional in globally supersymmetric
theories.

A spinorial version of the Bogomol'nyi bound has been derived previously
for the point-like solutions in three-dimensional supergravity
\cite{bbs,e1,e2}. However, that is not sufficient to prove the stability
of the $D$-term string solution in four dimensions. In this article we
shall reconsider the stability of $D$-term strings using the same
spinorial Witten-Nester method \cite{Witten,Nester}. A key step in
finding the original $D$-term string solutions was noting a fortuitous
cancellation between singular terms in the gravitino Killing spinor
equations \cite{dkvp,bbs,e1,e2}. The conical form of the metric ansatz
gives rise to singularities in the spin connection. However, these are
cancelled by equivalent contributions from the gauge field. It is this
same cancellation that allows one to derive the Witten-Nester form of the
Bogomol'nyi bound.

We begin in section 2 by reviewing the $D$-term string solution, with
particular attention being paid to the negative deficit angle ($\d<0$)
solution. Using a result from \cite{Comtet:1987wi}, we show that the
$\d<0$ string is not a regular solution to the field equations; it
necessarily violates the dominant energy condition, whereas the matter
Lagrangian does not. We argue that a $\d<0$ $D$-term string must have a
naked pathology, and therefore could not  exist as a counterexample to
any positive energy bound. In section 3, we review the Bogomol'nyi energy
functional approach used in \cite{dkvp} to define the energy density of
the $D$-term. We then discuss the more rigourous, Hamiltonian energy
density definition of Hawking and Horowitz \cite{hh}, which is naturally
associated to the linearised version of the Witten-Nester energy density.
In section 4, we will proceed to define an energy density integral for
string solutions in four-dimensions using a generalised Witten-Nestor
tensor. Using standard positive energy theorem techniques, we prove that
the Witten-Nester charge is manifestly non-negative. We then use a
linearised version of the surface integral expression to explicitly
calculate the energy density and Bogomol'nyi bound for the $D$-term
string solution. We find that the result agrees with the original
calculation, and we do not have to enforce an infrared cut-off to ensure
finiteness.

Finally, in section 4, we turn to a discussion of the stability of the
positive deficit angle ($\d>0$) $D$-term string solution. A key
assumption of Witten's proof of the positive energy theorem is that
matter obeys the dominant energy condition, which holds  for the
supergravity in question here. Assuming then that there are no internal
boundaries and that the generalised Witten condition holds, we argue that
the non-linear Bogomol'nyi bound derived from the Witten-Nester
expression implies that the $\d>0$ string is classically stable against
perturbations that asymptotically vanish at infinity. This can be seen as
a non-linear version of Gregory's {\it C-energy} argument for local
cosmic strings \cite{Gregory}. In particular, we find that the  $\d>0$
$D$-term string cannot decay perturbatively to the $\d<0$ string i.e.
that the $\d<0$ string is not a valid perturbation as it does not vanish
asymptotically. We show that any instanton that could provide a
non-perturbative, tunnelling process between supersymmetric solutions
with $\d>0$ and $\d<0$ would have to violate the dominant energy
condition, and therefore does not affect the positive energy proof. This
proves that the $\d>0$ $D$-term string of $\cN=1$ supergravity with
constant Fayet-Iliopoulos terms is stable.

\section{The $D$-term string in $\cN=1$ supergravity}

Let us begin by briefly reviewing the relevant aspects of four-dimensional $\cN=1$ supergravity with constant Fayet-Iliopoulos terms \cite{dkvp}. The Lagrangian for the bosonic sector of this theory is\footnote{We are using natural units, setting $M_P=1$.},
\begin{eqnarray}
e^{-1}{\cal L}=\frac{1}{2}R -\hat {\partial }_\mu
\phi\, \hat {\partial }^\mu \phi^*
 -\ft{1}{4} F_{\mu \nu } F^{\mu \nu }- V^D\,,
 \label{bosonic2}
\end{eqnarray}
where $\phi$ is the $U(1)$-charged Higgs field, the K{\"a}hler potential is given by $K=\phi^*\phi$ and the superpotential vanishes. The $D$-term  potential is defined by
\begin{equation}
V^D= \frac{1}{2} D^2 \qquad  D\, = g\xi - g \phi^*\, \phi \, ,
\label{Pphi}
\end{equation}
where $\xi$ is a constant that we choose to be positive. $W_\mu$ is an abelian gauge field,
\begin{equation}
F_{\mu\nu}\equiv {\partial }_\mu W_\nu- {\partial }_\nu W_\mu\, ,\qquad
\hat {\partial }_\mu \phi\equiv ({\partial }_\mu -\rmi g W_\mu) \phi\,.
\end{equation}
The fermions are Majorana spinors. However, it is often convenient to
split them into complex parts using left and right projectors: \be P_L =
\half(1+\g_5)~~~~~,~~~~~P_R = \half(1-\g_5)~. \ee The supersymmetry
transformations for the fermions (the Killing spinor equations) can then
be written as \be \delta \psi _{\mu}  = \hat\nabla_\mu \epsilon =
\nabla_\mu\epsilon + \frac{\rmi}{2} \g_5 A_\mu^B \epsilon\,,
\label{susy1} \ee \be \delta \chi_L =  \half(\not\! {\partial }-\rmi g
\not\! W) \phi \epsilon _R\,, \label{susy2} \ee \be \delta \lambda
=\frac{1}{4}\gamma ^{\mu \nu } F_{\mu \nu }\epsilon +\ft12\rmi \gamma _5
D
 \epsilon  \,. \label{susy3}
\ee
The covariant derivative on fermions is defined as $\nabla_\mu =  \partial _\mu  +\ft14 \omega _\mu
{}^{\ul{\a\b}}(e)\gamma _{\ul{\a\b}}$. The gravitino $U(1)$ connection  $A_\mu ^B$ plays an important role
in the gravitino transformations.
 \begin{eqnarray}
  A_\mu ^B&=&\frac{1}{2}\rmi\left[ \phi\partial _\mu \phi^* -\phi^* \partial _\mu \phi\right]
  + W_\mu  D\nonumber\\
  &=&\frac{1}{2}\rmi\left[ \phi\hat{\partial} _\mu \phi^* -\phi^* \hat{\partial} _\mu \phi\right]
  +g W_\mu  \xi  \,.
  \label{AmuBinphi}
\end{eqnarray}
The cosmic string solutions to this theory found in \cite{dkvp} solve the
Killing spinor equations \rf{susy1} - \rf{susy3} for some non-vanishing
$\epsilon$. The static ansatz for the metric in cylindrically symmetric
form is
\begin{equation}
  \rmd s^2= -\rmd t^2 +\rmd z^2+\rmd r^2 + C^2(r) \rmd \theta ^2\,,
 \label{tentativemetric}
\end{equation}
where the plane of the string is parametrised by $r$ and $\theta $. We
choose vierbein $e^1 =dr$ and $e^2 = C(r)d\theta$, which gives
$\o_\theta^{12} = -C'(r)$ as the only non-vanishing spin connection
component.

The Higgs field has the following form
\begin{equation}
\phi (r,\theta )\, = \, f(r)\,{\rm e}^{\rmi n \theta}\,,
 \label{stringhiggs}
\end{equation}
where $\theta$ is an azimuthal angle, and $f(r)$ is a real function that
outside the string core approaches the vacuum value $f^2=\xi$, for which
the $D$-term vanishes.  The gauge potential is given by
\begin{eqnarray}
 g W_\mu\, \rmd x^\mu = n\alpha (r) \,\rmd\theta \,
  \qquad \rightarrow \qquad
  F=\ft12F_{\mu \nu }\,\rmd x^\mu \,\rmd x^\nu =\frac{n \alpha '(r)}{g} \rmd
  r\,\rmd\theta  \,.
 \label{explicitW}
\end{eqnarray}
One can solve for the profile functions $\alpha(r)$ and $C(r)$ explicitly in limiting cases, and one sees that the metric describes a spacetime with a conical deficit angle proportional to the Fayet-Iliopoulos constant $\xi$. A globally well-behaved spinor parameter is defined by
\begin{equation}
  \epsilon _L(\theta) = \rme^{\mp\ft12\rmi\theta } \epsilon _{0L}\,,
 \label{epsilontheta}
\end{equation}
where $\epsilon _{0L}$ is a constant satisfying the following projection
\begin{equation}
  \gamma ^{12}\epsilon\, = \, \mp \rmi \gamma _5 \epsilon\,.
 \label{projectioneps12}
\end{equation}
By demanding that the following condition holds
\begin{equation}
  1-C'(r) = \pm A_\theta ^B\,,
 \label{diffeqrho}
\end{equation}
one can then find solutions to the gravitino Killing spinor equation:
\begin{equation}
\hat\nabla_\mu \epsilon_L = 0.
\end{equation}
As noted originally for three-dimensional supergravity \cite{bbs,e1,e2}, the key to solving this Killing spinor equation in a conical spacetime is the $U(1)$ charge of the gravitino. This allows the singular spin connection term to be cancelled precisely because both the $U(1)$ charge and the deficit angle are set by the Fayet-Iliopoulos term $\xi$.

When the distance $r$ from the string core is large, the solution \rf{tentativemetric} takes the form of an asymptotically locally flat conical metric with an angular deficit angle due to the constant FI term $\xi$:
\be
  \rmd s^2= -\rmd t^2 +\rmd z^2+\rmd r^2 + r^2\left(1\mp n\xi \right)^2 \rmd \theta ^2\label{metricFAR}~, \ee
with the composite gauge field given by $A_\theta ^B= n\xi $.  Note that in the limit $r \rightarrow\infty$   the full supersymmetry is restored as $F_{\mu\nu}=0$, $D=0$,  $\partial_r \phi= \hat\partial_\theta \phi= 0$ and $R_{\mu\nu}{}^{ab}=0$, which corresponds to  the enhancement of supersymmetry away from the core of the string. It is interesting to note that supersymmetry only fixes the metric function $C(r)$ up to a sign \cite{dkvp}, which will have some consequence for the definition of the string energy.

\bigskip

\subsubsection*{The negative deficit angle ($\d<0$) string}
A basic assumption in the proof of the positive energy theorem, which we
shall apply too in our proof of stability, is that matter satisfies the
dominant energy condition, which states that for any timelike vector $u^\mu$,
 $-T^{\nu}_{~\mu} u^\mu $ is non-spacelike. The dominant energy condition implies the weak, or null, energy condition, 
\be
T_{\mu\nu}u^\mu u^\nu \geq 0~,
\ee
where $u^\mu$ is now any non-spacelike vector. For a static spacetime, such as that of the $D$-term string, an equivalent statement is that $T_{00}$
dominates over any other stress-energy tensor component in any
orthonormal frame \cite{HE}: \be T_{00} \geq |T_{ij}| ~~~~~\mathrm{for~
any}~ i,j~, \ee where $i,j$ are purely spatial indices. Physically this
corresponds to the principle of causality; that matter energy flows at
subluminal speeds. In particular, we see that the dominant energy
condition implies that matter energy density is strictly non-negative in
any orthonormal frame. It is well known that regular matter, such as
massless scalar and vector fields, satisfies the dominant energy
condition, however, a violation of this condition could occur if a
massive scalar field had a sufficiently negative potential. It is
straightforward to check that the supersymmetric Lagrangian
\rf{bosonic2}, with its manifestly positive potential, satisfies this
condition.

We shall now argue that the static string asymptotic solution with
negative deficit angle (i.e. with $C(r) = r (1+|n| \xi)$) should violate
the assumption of the dominant energy condition when analytically
continued over the whole space. As the matter Lagrangian does not, it
cannot be a full solution unless we add other matter. Therefore, it does
not affect our arguments relating to the positive energy theorem. In
contrast, the $\d>0$ D-term string does not require the dominant energy
condition to be violated and the solution is regular throughout
\cite{dkvp}.

One of the requirements for the Witten-Nester positive energy is that the
Einstein equations be solved throughout the spacelike surface on which
one wishes to compute the energy of the solution. As was shown in
\cite{Comtet:1987wi}, a general string solution can only have a $\d<0$ if
its matter sector has $T_{0 0} < 0$, i.e. it violates the dominant energy
condition. We shall now summarise the basic ingredients of this argument:
First, choose a general three-dimensional, cylindrically symmetric,
spacelike Cauchy surface $\Sigma_3$. Then write the Einstein equations in
terms of the Ricci scalar $^3 R$ of $\Sigma_3$ (i.e. the `initial value
constraint') using the Gauss-Codazzi relations. It is possible to rewrite
$^3 R$ in terms of the trace of the extrinsic curvature  $^2K$ of a
two-dimensional submanifold $\Sigma_2$ transverse to the string, plus
terms that are manifestly positive. Integrating the initial value
constraint over this two-dimensional submanifold and applying the
Gauss-Bonnet theorem, one obtains a relation of the form:
\begin{equation}
\delta \sim +\int_{\Sigma_2} T_{00} + ( \ldots )^2\,,
\end{equation}
where $\d$ is the deficit angle. Hence, we see that it is only possible to have a solution with $\delta < 0$ if the Lagrangian violates the dominant energy condition. However, one can easily check that the Lagrangian of our model \eqref{bosonic2} can only have $T_{0 0} \geq 0$. Assuming a static ansatz, with a vanishing timelike component for the composite gauge field, $A^B_0=0$, our system \eqref{bosonic2} has the following $T_{00}$:
\begin{equation}
T_{00} = (-g_{00})\,\Big(\hat{\partial}_i\phi\,\hat{\partial}^i\phi^* + \tfrac{1}{4}\,F_{ij}F^{ij} + \tfrac{1}{2}\, D^2\Big)\,.
\end{equation}

 This means that the $\d<0$ $D$-term string fails to solve the Einstein equations in some region of the putative Cauchy surface\footnote{We thank G. Gibbons and S. Ross for useful discussions on this point.}, and therefore needs a source with negative $T_{0 0}$. Although the $\delta < 0$ solution is not known in closed form for small radius, the string ansatz we are using \eqref{tentativemetric} does not have a $g_{t t}$ component, and hence does not allow it to have horizons in the interior of the solution. This means that this defect, which requires the presence of a source, sweeps out a worldvolume over infinite time.  To see this, one can roughly think of the defect as a region defined by $r<r_d$ for some $r_d$. Then, the absence of horizons in the metric implies that the `$r$' coordinate is everywhere spacelike, so the defect is present at all times. In other words, the region of the solution that violates Einstein's equations is naked, which means that no spacelike surface will be able to avoid it. Therefore, the positive energy theorem does not apply to the $\d<0$ solution.

This is reminiscent of the Schwarzschild solution with negative mass.
Although the positive mass Schwarzschild solution fails to solve the
Einstein equations at the curvature singularity, the Kruskal extension
shows that the latter is actually a spacelike region. This means that it
is possible to choose a spacelike Cauchy surface that avoids the
singularity, thereby guaranteeing that the equations of motion are solved
throughout it. The negative mass solution, however, has a naked
singularity that cannot be avoided by any spacelike surface. This
excludes it from the Witten-Nester positive energy theorem \cite{ghhp}.

\noindent Throughout the rest of this paper, we will focus on the $\d>0$ solution and will return to the $\d<0$ case only in the discussion of the stability of the positive case.

Before proceeding, we shall comment briefly on the possibility of a
rotating solution, i.e., a $D$-term string with angular momentum. In
\cite{Deser:1983tn}, a spatially localised spinning solution of
three-dimensional gravity was found. This stationary solution is the
three-dimensional analogue of the Kerr solution, with the angular
momentum being determined by the momentum densities $T_{0i}$. However,
the three-dimensional spinning solution can have vanishing energy density
($T_{00}=0$) and non-zero angular momentum, and therefore obviously
violates the dominant energy condition. Indeed it was already noted in
\cite{Deser:1983tn} that such spacetimes are not causal. To our
knowledge, an analogous spinning string solution of this form has not
been found in four-dimensional $\cN=1 $ supergravity with a constant
Fayet-Iliopoulos term, and it would be interesting to understand whether
it exists. While it is difficult to discuss the properties of such a
hypothetical solution, it would seem reasonable to assume that it would
suffer from the same causality problem as its three-dimensional
counterpart, and thus it would also be excluded from the Witten-Nester
positive energy theorem.

\bigskip

\section{Defining the energy of the $D$-term string}

In order to address the stability of the $D$-term string solution, it
shall be useful to first reconsider its energy definition. Various
methods exist for defining energy in spacetimes with non-trivial
asymptotic structure.  In \cite{dkvp}, the string energy density was
defined using a Bogomol'nyi style argument. As the solution is
time-independent, the ansatz could be directly inserted into the action
with Gibbons-Hawking boundary terms included to give an energy
functional. The integral was then restricted to only run over directions
transverse to the string to ensure it produced a finite result. Using the
Bogomol'nyi method, this integral was then written in the following way
\begin{eqnarray}
{\cal \mu}_{\rm string}&=& \int \, \rmd r\rmd\theta\, C(r)\left\{  \,
 |(\hat{\partial}_r \phi  \, \pm \, {\rm i}C^{-1} \, \hat{\partial }_\theta ) \phi|^2 \, + \,
{\fft12}\left[ F_{12} \, \mp   D
\right]^2 \right\}  \,+\label{ebpsstring}\\
&+& \int \rmd r \rmd\theta \,  \left[\partial _r\left( C'\pm
A_\theta\right) ^B\mp\partial _\theta A_r^B\right]-\left.\int
\rmd\theta\, C'\right|_{r=\infty }+\left.\int \rmd\theta\,
C'\right|_{r=0}~. \nonumber
\end{eqnarray}
The condition arising from the gravitino Killing spinor equation~(\ref{diffeqrho}) implies that the first term in the second line in (\ref{ebpsstring}) vanishes. The first line vanishes by the remaining Killing spinor equations $\d \l = 0 = \d\chi_L$. The energy density is thus given by the difference between the boundary terms at $r=0$ and at $r =\infty$ \cite{dkvp}:
\begin{equation}
  \mu _{\rm string}=2\pi \left(\left.C'\right|_{r=0 }-\left.C'\right|_{r=\infty }\right)= \pm 2\pi n \xi \, ,
\label{energy}
\end{equation}
which agrees with the expected answer for a cosmic string solution \cite{Thorne,book}.

\subsubsection*{A Hamiltonian energy definition}

Let us now reformulate the energy definition for the $D$-term string
using a more rigourous approach. We shall follow the approach of Hawking
and Horowitz \cite{hh}, who have proposed a counterterm subtraction
method to define the energy of a non-compact spacetime from its
Hamiltonian.  The counterterm here is nothing more than the Hamiltonian
of an appropriately identified reference, or background, spacetime
denoted $\mathcal{H}_0$. One performs a canonical ADM decomposition\footnote{A specific analysis of the ADM decomposition in
cylindrically symmetric spacetimes has been given by Frolov et al
\cite{Frolov:1989er}.} of
the non-compact spacetime into $\mathbb{R}\times \Sigma_t$ and defines
its total energy as the value of the physical Hamiltonian, which itself
is defined by:
 \be
\mathcal{H}_{phys} \equiv \mathcal{H} - \mathcal{H}_0 = \int_{\Sigma_t}
\,[NH+N_iH^i] -  \int_{S^\infty_t} \left(N ({}^2\!K - {}^2\!K_0)  -
N^ip_{ij}r^j  \right)~, \label{hh}
 \ee
where the background contributes just a boundary term,
 \be \mathcal{H}_0
= - \int_{S^\infty_t}\sqrt {h_0} N ~{}^2\!K_0~.
 \ee
Here $N$ is the lapse function, $N^i$ is the shift function, and $H$ and
$H^i$ are the constraints.  Indices are raised and lowered by $g_{ij}$,
the intrinsic metric on the spatial hypersurfaces $\Sigma_t$. The
conjugate momenta $p_{ij}$ and shift function $N^i$ vanish for the
$D$-term string solution, as does the Gauss constraint, which we have not
written explicitly. The boundary terms are written in terms of the
extrinsic curvature ${}^2K$ of a 2-surface $S^\infty_t$ in $\Sigma_t$ ,
where $S^\infty_t$ is formally a family of 2-surfaces with metric
$h_{ij}$, defined by the intersection of $\Sigma_t$ with the asymptotic
boundary $\Sigma^\infty$.

The background Hamiltonian is defined solely by ${}^2K_0$, the embedding
of the 2-surface into the spatial 3-slice of the appropriately identified
background spacetime. In order to evaluate the energy of a particular
cosmic string solution, it was then argued in \cite{hh} that the
appropriate background spacetime should be the string metric with
vanishing deficit angle, i.e. Minkowski space. It is straightforward to
calculate the extrinsic curvature traces,
 \be
 \sqrt {h}~{}^2\!K = -(1\mp n\xi) ~~~~~,~~~~~~\sqrt {h_0}~{}^2\!K_0 = -1~.
  \ee
For the case of the $D$-term string $N=1$, and the generic surface
$S^\infty_t$ is the cylinder $\mathbb{R}_z \times S^1_{\theta}$. Any
integration over the string worldvolume direction $\mathbb{R}_z $ will
produce an infinite contribution, thus we should regulate the integral
\rf{hh} to have a finite result. This can be formally achieved by
wrapping the string on a circle $S^1_z$ of fixed radius $R$
\cite{ght,ks,tz}. The string energy per unit length is  then defined by
 \be
  E = \frac{1}{2\pi R} \int_{S^1_z \times S^1_{\theta}} N(\sqrt
{h}~{}^2\!K - \sqrt {h_0}~{}^2\!K_0) = - \int d\theta (\mp n\xi) = \pm
2\pi n \xi~, \label{mhh}
 \ee
which agrees with the known result for cylindrically symmetric spacetimes
\cite{Thorne,book}, and thus  also the Bogomol'nyi approach. The
advantage of this definition is that one does not have to use exceptional
symmetries of the spacetime in question in order to define the energy in
general. Also, while it was necessary to wrap the spatial component of
the string worldvolume in order to have a finite result, we do not have
to regulate the integral otherwise. One may have anticipated the need for
an infrared cut-off in such a conical spacetime. However, the boundary term $\sqrt{h}~{}^2\!K$ is itself already finite at large $r$. Finally, we will see that this form of the energy definition appears naturally in
the linearised version of the spinorial analysis, to which we now turn.

\section{Positive energy and stability}

The energy definitions given in the previous section have concrete
physical interpretations. However, both are problematic if one wants to
consider the perturbative stability of the $D$-term string background.
The Bogomol'nyi bound argument applied to the $D$-term string is
certainly indicative of its stability. Unfortunately, by directly
substituting the metric ansatz into the action one is limiting any
perturbations to lie within the same symmetry class as the background,
hence no satisfactory stability argument can be given by linearising the
BPS equations. One can consider a general linearised perturbation
analysis of the Hamiltonian energy, but this generically leads to terms
of indefinite sign, even around flat spacetime \cite{Brill}. The most
rigourous way to prove the stability of a solution in a generally
covariant theory is to use the spinorial Witten-Nester technique
\cite{Witten,Nester}. One defines the Witten-Nester four-momentum as a
surface integral, one component of which is the total gravitational
energy i.e. the ADM-like mass plus a charge contribution. Using Gauss'
law to rewrite this as a volume integral, it is then straightforward to
show that the total energy is non-negative, and only vanishes for
supersymmetric solutions. This implies that any perturbation around a
supersymmetric background solution must contribute positively to the
total energy, and therefore cannot cause an instability. We will now
apply this method to the static $D$-term string, where it works in much
the same way as for supersymmetric solutions to three-dimensional
supergravity \cite{bbs,e1,e2}. Once again, we will only consider $D$-term
strings with positive deficit angle $\d>0$. We shall argue that they are
non-perturbatively stable when considered as solutions of the
supergravity theory and in particular, that they cannot decay to the
string with negative deficit angle $\d<0$.

\subsubsection*{Witten-Nester energy and the Bogomol'nyi bound}

We begin by defining the generalised Witten-Nester 2-form $ E =
\frac{1}{2} E_{\mu\nu} dx^\mu\wedge dx^\nu$ \cite{ghhp,gh,ghw,Boucher,Townsend,fg}:
\be
E^{\mu\nu} = \bar\eta \,\gamma^{\mu\nu\rho} \hat\nabla_\rho \eta~,
\ee
where we are using the supercovariant derivative defined by the gravitino supersymmetry transformation \rf{susy1}, and $\eta$ denotes a commuting spinor function that asymptotically tends to a background Killing spinor $\hat\nabla_\rho \eta =0$.

We now define the Witten-Nester four-momentum as the integral of the dual of $E$
\begin{eqnarray}
P_\mu v^\mu \!\!&=&\!\!  \int_{\pt M} \ast E \\ \nn \\ \!\!&=&\!\! \frac{1}{2}\int_{\pt M} dS_{\mu\nu} E^{\mu\nu}  = \int_{M} d\Sigma_\nu \nabla_\mu E^{\mu\nu}~,
\end{eqnarray}
where $v^\mu = \bar\eta \g^\mu  \eta$. In the second line we have assumed
that there are no internal boundaries since the $\d>0$ D-term is regular
\cite{dkvp}, and used Gauss' law to write a volume integral. At this
point, one should understand that ${\pt M}$ is the two-dimensional
boundary of an arbitrary three-dimensional subsurface $M$. Before proving
that the energy $P_0$, defined by the Witten-Nester integral is positive,
let us first express it in more familiar terms. In order to identify the
quantities appearing in this boundary expression, we need only consider
linearised perturbations in the integrand. We can then manipulate the
surface integral expression in the standard way to find
\begin{eqnarray}
P_\mu v^\mu &=& \half \int_{\pt M} dS_{\mu\nu} \bar\eta \,\gamma^{\mu\nu\rho} \hat\nabla_\rho \eta \nn \\
&=& \frac{1}{8}\int_{\pt M} dS_{\mu\nu} \varepsilon^{\mu\nu\rho\sigma}\varepsilon_{\d\a\b\s} \Delta\o_\rho^{\ul{\a\b}}e_{\ul \a}^\a e_{\ul \b}^\b\,\bar\eta \g^\d  \eta - \frac{1}{4}\int_{\pt M} dS_{\mu\nu} \varepsilon^{\mu\nu\rho\sigma} A_\rho^B \bar\eta \g_\s  \eta~. \label{lhsew} 
\end{eqnarray}
The first term in \rf{lhsew}, which we shall denote $ {\cal P}_\m v^\m$, is Nester's expression for the gravitational four-momentum \cite{Nester}, where $\Delta\o_\rho^{\ul{\a\b}}$ is the difference of the spin connection with respect to the reference spacetime with $A_\rho^B =0$, i.e. Minkowski spacetime.  The second term in \rf{lhsew}, which we shall denote $J_\m^Rv^\m$, defines the R-charge of the string, i.e. the holonomy of composite gauge potential. We have assumed here that perturbations fall-off sufficiently quickly such that the integral is well-defined. Once again, this surface charge integral must be regulated  i.e. we have to wrap the spatial worldvolume of the string such that ${\pt M} =\mathbb{R}_z \times S^1_{\theta} \rightarrow S^1_z \times S^1_{\theta}$, and then integrate out the $z$-contribution. The charge integral is then defined only over spatial directions transverse to the string, and it is formally the same as the equivalent three-dimensional expression \cite{bbs,e1,e2,Bombelli,ght}. Thus, the fall-off conditions required for three-dimensional metric perturbations to produce finite contributions to boundary integrals, which have been described in \cite{Marolf}, also apply here. In order to calculate this linearised charge integral for an arbitrary string configuration, one should also ensure that no Kaluza-Klein type charges appear in the boundary energy density integral in three dimensions \cite{Bombelli,ght}.

Let us now use \rf{lhsew} to evaluate the total gravitational energy of the $D$-term string solution itself. As discussed above, for the string solution we must regulate the worldvolume integral in the charge definition to ensure it produces a finite result. Thus we choose to wrap the spatial worldvolume direction of the string on an $S^1$ of fixed radius $R$ and calculate the corresponding energy density. If we now fix coordinates such that $\Sigma$ has a simple timelike normal and choose $v^\mu$ to be the asymptotic timelike Killing vector of the spacetime,  we can insert the vierbein and spin connection of the $D$-term string metric \rf{metricFAR} into \rf{lhsew} to find \begin{eqnarray}
P_\mu v^\mu &=& \frac{1}{2 \pi R}\int dz d\theta \left(\left[1- C'(r)\right] ~ \mp ~ A^B_\theta \right) v^0 ~.\label{lhsew-dts}
\end{eqnarray}
where the $\mp$ comes from using the projection condition
\rf{projectioneps12}. We can now explicitly identify the R-charge $Q^R$
of the $D$-term string,
 \be
Q^{R} = J^R_0 =  \pm \frac{1}{2 \pi R} \int dz d\theta A^B_\theta v^0~.
 \ee
We see that the first term in \rf{lhsew-dts}, which is equivalent of the
ADM mass, produces the correct result for the string, in agreement with
both the Hamiltonian and Bogomol'nyi style arguments. In fact, it is
known that the Nester definition of energy that appears here is one of an
equivalent set of boundary term energy definitions \cite{in}. These
equivalent definitions all arise as rewritings of the Gibbons-Hawking
term, and therefore are formally the same as the Hawking-Horowitz
expression. It is in this sense that the Hawking-Horowitz energy is
naturally related to the spinorial definition. Moreover, both the
Witten-Nester and Hawking-Horowitz energies are computed without having
to insert a cut-off at large $r$ i.e. once the spatial worldvolume
contribution is accounted for, both charges are well-defined and produce
a finite result.


In order to prove the positivity of the Witten-Nester four-momentum we now  turn to the volume integral expression
\be
P_\mu v^\mu = \int_{M} d\Sigma_\nu \nabla_\mu E^{\mu\nu}~.
 \ee
We would like to show that the right-hand side of this expression is
positive by rewriting it as a sum of squares. We are no longer interested
in calculating the energy of a particular string solution. Rather, we
want to show that an arbitrary on-shell perturbation of a supersymmetric
solution that vanishes asymptotically, but is otherwise unbounded,
contributes a positive amount to the total energy. As such, we shall not
wrap the spatial direction of the string worldvolume, such that $M$ is a
two-dimensional region, but consider the full three-dimensional volume
integral with $M=\mathbb{R}_r \times \mathbb{R}_z \times S^1_{\theta} $,
allowing for the most general perturbations. A lengthy calculation using
the standard manipulations \cite{gh,ghw,Boucher,Townsend,fg} then leads
to the following expression
 \be
P_\mu v^\mu = \int_{M} d\Sigma_\nu\left( \overline{\hat\nabla_\mu\eta}\g^{\mu\nu\rho}\hat\nabla_\rho\eta + \overline{\d\l} \g^\nu  \d\l + 2 \overline{\d\chi_L} \g^\nu  \d\chi_L + 2 \overline{\d\chi_R} \g^\nu  \d\chi_R\right)~,
 \ee
where $\d\l$ and $\d\chi_{L,R}$ are the supersymmetry transformations
\rf{susy1}-\rf{susy3}, defined now with a commuting spinor parameter
$\eta$. Choosing $\Sigma$ to be an initial hypersurface with simple
timelike norm, we find that the Witten-Nester charge becomes
 \be
P_\mu v^\mu = \int_{M} d\Sigma_0 \left( - \hat\nabla_i\eta^\dagger \g^{i}\g^j \hat\nabla_j \eta +  \eta^{ij}\hat\nabla_i\eta^\dagger \hat\nabla_j \eta  + \d\l^\dagger  \d\l +  2\d\chi_L ^\dagger  \d\chi_L +  2\d\chi_R^\dagger  \d\chi_R\right)~.
 \ee
We now choose spinors that obey the generalised Witten condition,
 \be
\g^j\hat\nabla_j\eta = 0~,
 \ee
which implies that the integral is manifestly positive:
 \be
P_\mu v^\mu = \int_{M} d\Sigma_0 \left( \eta^{ij}\hat\nabla_i\eta^\dagger \hat\nabla_j\eta + \d\l^\dagger  \d\l +  2\d\chi_L ^\dagger  \d\chi_L + 2 \d\chi_R^\dagger  \d\chi_R\right)~. \label{rhsew}
 \ee
In order that this implies a positive energy, we must show the Killing
vector $v^\m$ is non-spacelike and future directed. In fact, as we are
using commuting Majorana spinors it is straightforward to apply the Fierz
identity to $\bar\eta \g^\m \eta \bar\eta \g_\m  \eta$ to check that
$v^{\m}$ is null. As our choice of initial hypersurface $\Sigma$ was
arbitrary, we can allow for arbitrary variations of it. This means that
our expressions get promoted to fully covariant versions, and the Witten
condition becomes $\g^\m\hat\nabla_\m\eta =0$. If we now use the
covariant form of our result \rf{rhsew} in conjunction with the
expression for the Witten-Nester four-momentum \rf{lhsew}, we reproduce
the Bogomol'nyi bound for the $D$-term string:
 \be
\overline\eta_0\left({\cal P}_\nu- J^R_\nu \right)\g^\n\eta_0 \geq0~. \label{ewbb}
\ee
Looking again at \rf{rhsew}, we see that this inequality is saturated when the solution is supersymmetric, i.e. when $\d\l=\d\chi=\d\psi_\mu=0$. Here $\d\psi_i$ has been promoted to $\d\psi_\mu$ by allowing for arbitrary variations of the hypersurface $\Sigma$. It is possible to bring the Bogomol'nyi bound \rf{ewbb} into the more familiar form ${\cal P}_0 - Q^R \geq 0$ by taking the trace over the basis of spinors.

\subsubsection*{Stability of the $\d>0$ $D$-term string}

We shall now discuss to what extent the bound \rf{ewbb} implies stability
for the positive deficit angle solutions. Let us begin by discussing how
this result differs from the three-dimensional bound derived in
\cite{bbs,e1,e2}. We know that in order to have a finite result for the
energy of a $p$-brane in D dimensions, one must wrap its spatial
worldvolume on a $p$-torus, and consider the corresponding energy density
\cite{ght,ks,tz}. As discussed above, one must also ensure that all
perturbations fall-off sufficiently quickly such that no Kaluza-Klein
type charges appear in the boundary energy density integral in (D$-p$)
dimensions.  For the $D$-term string this amounted to wrapping the
$z$-direction on an $S^1$ of fixed radius and calculating the
corresponding energy density in three dimensions. If no Kaluza-Klein type
charges appear, then the energy density calculation for a particular
solution will be nothing more than the corresponding three-dimensional
version \cite{bbs,e1,e2}. In this sense the $D$-term string is seen as a
solution of a consistent reduction to the massless sector of the
resulting three-dimensional supergravity theory. However, if one wishes
to consider the stability of this class of string solutions, it is the
volume integral in four dimensions that must be studied. The
perturbations that would source Kaluza-Klein charges in three dimensions
may be asymptotically small, and thus not contribute to the boundary
expression, but they are unbounded in the bulk of the spacetime.
Experience with string-like solutions in higher dimensions tells us that
possible instabilities would arise in the massive Kaluza-Klein tensor
perturbations in the dimensionally reduced theory \cite{Gregory:1993vy}.
Thus it is not sufficient to prove stability by studying just the
massless sector in three dimensions; rather one must consider the full
Kaluza-Klein tower, or equivalently the original four-dimensional theory.

Having understood that it is necessary to reconsider the stability
question from the four-dimensional perspective, one can easily see that
the above proof of the Bogomol'nyi bound proves the perturbative
stability of the $D$-term string with metric \rf{metricFAR} to all order
in perturbations. While we have not stated the explicit fall-off
conditions for such perturbations (see \cite{Marolf} for details), the
primary constraint is that they vanish asymptotically. It is not
difficult to check that a negative deficit angle string viewed as a
metric perturbation of the positive deficit angle string does not vanish
asymptotically, therefore is not a perturbative decay channel.
Nevertheless one should also question whether non-perturbative decay
channels exist.

When studying non-perturbative quantum tunnelling effects, one must relax
the preservation of boundary conditions by the decay process in question
and allow for processes that can alter them dynamically.  The
archetypical example of this is the Coleman-de Luccia bounce solution in
\cite{Coleman:1980aw}. In that paper a mechanism is described by which
Minkowski spacetime, viewed as the false vacuum of a certain theory, can
decay into AdS, viewed as the true vacuum, by forming a bubble of true
vacuum at the origin of spacetime that grows at the speed of light,
quickly engulfing the universe. If we do not want to exclude such an
interesting non-perturbative effect in our case we must be willing to
relinquish the fixed asymptotic deficit angle and allow for a decay
channel that can change it. However, Taylor has argued in
\cite{Taylor-Robinson:1996fk} that decay modes via bubble nucleation are
inconsistent with ten and eleven dimensional supergravity
theories\footnote{Note that it is possible to find instantons describing
the decay of non-supersymmetric strings via black hole pair production
\cite{Hawking:1995zn,Eardley:1995au}.}. The arguments, which we will now
sketch, carry over to other supergravity theories.

The decay of a false vacuum into a true vacuum in the semi-classical
theory is described in two steps. First, one finds a `bounce' solution to
the Euclidean equations of motion that asymptotes to the false vacuum,
but is allowed (expected) to be different from it in the interior. This
describes the nucleation of a `bubble of true vacuum' inside a universe
in the false vacuum through barrier penetration. On this solution, one
must find a zero-momentum hypersurface (i.e. surface of zero extrinsic
curvature w.r.t. Euclidean time) from which one can obtain a Lorentzian
solution via Wick rotation\footnote{Unless $g_{0i}$ metric components
vanish, the Lorentzian metric will not be real
\cite{Taylor-Robinson:1996fk}. Thus any `rotating' Euclidean solutions
are ruled out as possible instantonic decay channels.}. This describes the
evolution of the bubble in time. Such a zero-momentum hypersurface can be
seen as a time-independent spacelike hypersurface in the Lorentzian
theory. If this solution is required to asymptote to the original false
vacuum solution, then it must also admit an asymptotically Killing spinor
that is well-defined on the whole hypersurface. However, the positive
energy theorem implies that the energy of this solution can only be
higher than that of the false vacuum, making it energetically
unfavourable for the nucleation to take place. If the energy is equal to
that of the false vacuum then the spinor must be globally Killing, which
means the solution is no different from the false vacuum solution. In
other words, no bubble is being nucleated.

This argument substantiates our claim that the positive energy proof is non-perturbative for the $\d>0$ $D$-term string. Realistically, however, the supergravity model we study should be viewed as being embedded in a larger model, in which one cannot exclude other non-perturbative decay channels such as monopole creation.

\section{Discussions}

We have reconsidered the energy and stability of the $D$-term string
solution of $\cN=1$ supergravity with constant Fayet-Iliopoulos terms.
Our method was to use the Witten-Nester approach to prove that the
positive deficit angle ($\d>0$) supersymmetric $D$-term string was
stable. Using the gravitino supersymmetry transformation as a guide, we
defined the generalised Witten-Nester 2-form and the associated surface
and volume form of the charge integrals. We were then able to rewrite the
volume integral as a sum of squares plus one term of indefinite sign
using standard manipulations. The important feature of the $D$-term
string solution is the precise cancellation between the singular
components of the spin connection and the holonomy of the composite gauge
field $A_\mu^B$, which allowed the Killing spinor equations to be solved
in the asymptotically conical background \cite{dkvp,bbs,e1,e2}.  This
same cancellation implies we can enforce a generalised Witten condition
on the commuting spinor parameters, $\g^i\hat\nabla_i\eta =0$. This
allows us to consistently remove the indefinite term in the charge
integral, thus proving that the Witten-Nester expression for the total
gravitational energy is bounded from below.

By considering the surface integral energy expression, we were able to reproduce the known results for the $D$-term string energy density, R-charge and the Bogomol'nyi bound without the need for an infrared cut-off. We also argued that the spinorial expression for the string energy density, written in the canonical Nester form, is formally equivalent to the Hawking-Horowitz version of the Hamiltonian energy definition, which we presented as an alternative to the Bogomol'nyi style approach advocated in \cite{dkvp}.

The key step in proving the stability of the $\d>0$ $D$-term string is to
show that the $\d<0$ string does not stand as counterexample to the
positive energy theorem for this class of supersymmetric solutions. We
have seen that the $\d<0$ $D$-term string is not a proper solution, as it violates the dominant energy condition whereas the matter
Lagrangian does not. This implies that the $\d<0$ string has a naked
pathology  (i.e. a  region, not masked by a horizon, in which the
solution does not solve the field equations), and therefore it is not a
counterexample to the positive energy theorem. We also show that the
$\d<0$ string is not a viable perturbative or non-perturbative decay
channel for the $\d>0$ string. Together with the Bogomol'nyi bound
derived using the Witten-Nester techniques, this implies that the $\d>0$
$D$-term string of $\cN=1$ supergravity with constant Fayet-Iliopoulos
term is stable. At the level of perturbative stability this can be seen
as an extension of Gregory's analysis of the linearised stability of
local cosmic strings \cite{Gregory}. The main advantage of the spinorial
stability proof we present here is that one is forced to clarify the role
that the $\d<0$ string, a point that is often overlooked.

The obvious limitation of our analysis is that it does not cover possible
instabilities that arise when the $\cN=1$ supergravity with
Fayet-Iliopoulos terms is embedded into some grand unified model. In that
case it is well known that other decay channels arise, such as local
strings decaying to monopoles along their worldvolume via the Schwinger
process \cite{cmp}. It would be interesting to extend our arguments to
cases with more complicated matter sectors, which may provide an insight
into the breakdown of the positive energy theorem in those cases.  Also,
it is clearly important to have a better understanding of the behaviour
of a local cosmic string with a negative deficit angle $\d<0$, and in
particular to assess whether a more general metric ansatz would allow a
$\d<0$ string to have a horizon. While this is unlikely, it is a crucial
aspect of these string models and should certainly be considered more
thoroughly.

\section*{Acknowledgments}

We would like to thank Ana Ach\'{u}carro, Jos{\'e} Edelstein, Dan Freedman,
Gary Gibbons, Ulf Gran, Luca Martucci and Joris Van den Bergh for useful
discussions. PS would like to thank the members of the C.P.T. Durham,
where this work has been presented previously, and especially Simon Ross
for comments and suggestions. This work is supported in part by the
Federal Office for Scientific, Technical and Cultural Affairs through the
``Interuniversity Attraction Poles Programme -- Belgian Science Policy"
P5/27 and by the European Community's Human Potential Programme under
contract MRTN-CT-2004-005104 ``Constituents, fundamental forces and
symmetries of the universe''.

\subsection*{Conventions}

Our conventions shall follow \cite{tools}. We use a `mostly plus' metric.
Greek indices are four-dimensional, and where necessary Latin indices
will denote purely spatial directions. Flat indices will always be
underlined.


Our gamma matrices satisfy $\{\g_\mu,\g_\n\} = 2g_{\m\n}$. A barred
spinor is Majorana conjugate, and we define the Majorana condition by
$\bar\l= \l C = \l \g_0\g_2 =  -\rmi \l^\dagger \g_0$ for an
anti-commuting spinor, and $\bar\eta =  \eta^\dagger \g_0$ for a
commuting spinor. $\g_0$ is anti-Hermitian, while $\g_5 = \rmi
\g_0\g_1\g_2\g_3$ and $\g_i~(i=1,2,3)$ are Hermitian. For chiral spinors,
e.g. $\bar\l \equiv P_L\l$, this implies $\bar\l_L = -\rmi
(\l_R)^\dagger\g_0$. Some useful gamma matrix identities are
\begin{equation}
\g_5\g_{ \ul \a_1} = \frac{\rmi}{3!} \ep_{\ul \a_1\ldots  \ul \a_4}\g^{\ul \a_2\ul \a_3 \ul \a_4}~~~~~,~~~~~ \rmi \ep^{\ul \a_1\ldots  \ul \a_4}\g_{\ul \a_4}  = \g^{\ul \a_1\ul \a_2 \ul \a_3}\g_5~.
\end{equation}
In our conventions, the commutator of two covariant derivatives on a spinor acts as follows:
\begin{equation}
[\nabla_{\mu}, \nabla_{\nu}]\, \eta = \tfrac{1}{4}\,{R_{\mu \nu}}^{\ul{\a\b}}\,\gamma_{\ul{\a\b}}\, \eta\,.
\end{equation}
It is useful to note the following when proving the Bogomol'nyi bound from the Witten-Nester charge:

\begin{eqnarray}
\nabla_\mu E^{\mu\nu} &=& \bar\eta\g^{\mu\nu\rho}\hat\nabla_\mu\hat\nabla_\rho\eta + \overline{\hat\nabla_\mu\eta}\g^{\mu\nu\rho}\hat\nabla_\rho\eta~,
\end{eqnarray}
\begin{eqnarray}
\bar\eta\g^{\mu\nu\rho}\hat\nabla_\mu\hat\nabla_\rho\eta &=& \half \bar\eta G^\nu_\mu \g^\mu \eta - \frac{i}{4}\bar\eta \g^{\mu\nu\rho} F_{\mu\rho}\g_5 \eta - \frac{1}{2}\bar\eta \g^{\mu\nu\rho} \hat\pt_\mu \phi \hat\pt_\rho \phi^*\g_5 \eta \label{ac1}~,  \label{ac2}
\end{eqnarray}
where $G_{\m\n}$ is the Einstein tensor.

\end{document}